%% file: main.tex
\documentclass[conference]{IEEEtran}
\usepackage[T1]{fontenc} % Use T1 encoding for special characters
\usepackage{amsthm}
\usepackage{times}
\usepackage{breqn}
\usepackage{algorithm}
\usepackage{algpseudocode}
\algrenewcommand\algorithmicrequire{\textbf{Input:}}
\algrenewcommand\algorithmicensure{\textbf{Output:}}

\usepackage{epsfig}
\usepackage{epstopdf}
\usepackage{subcaption}
\usepackage{multirow}
\usepackage{comment}
\usepackage{listings}
\usepackage{xcolor}
\usepackage{soul}
\usepackage{xspace}
\usepackage{url}
\usepackage{booktabs}
\usepackage{hyperref}

\newcommand{\bigO}[1]{$\mathcal{O}\paren{#1}$\xspace}

\newcommand{\paren}[1]{\left(  #1 \right)}
\usepackage{cite}
\usepackage{amsmath,amssymb,amsfonts}
\usepackage{graphicx}
\usepackage{textcomp}
\usepackage{xcolor}

\def\BibTeX{{\rm B\kern-.05em{\sc i\kern-.025em b}\kern-.08em
    T\kern-.1667em\lower.7ex\hbox{E}\kern-.125emX}}

\makeatletter
\newcommand{\linebreakand}{%
  \end{@IEEEauthorhalign}
  \hfill\mbox{}\par
  \mbox{}\hfill\begin{@IEEEauthorhalign}
}
\makeatother

\begin{document}

\title{Designing Parallel Algorithms for Community Detection using Arachne}
\author{

\IEEEauthorblockN{Fuhuan Li, Zhihui Du, David A. Bader}
\IEEEauthorblockA{\textit{Department of Data Science} \\
\textit{New Jersey Institute of Technology}\\
Newark, New Jersey, USA\\
\{fl28,zhihui.du,bader\}@njit.edu}

%\and
%\linebreakand
}

\maketitle

%\begingroup\renewcommand\thefootnote{$^*$}
%\footnotetext{This research was partially supported by NSF grant number CCF-2109988.}
%\footnotetext{This research was partially supported by NSF grant number XXXXXXXXXXX.}
%\endgroup
\begin{abstract}
%\looseness=-1
The rise of graph data in various fields calls for efficient and scalable community detection algorithms. In this paper, we present parallel implementations of two widely used algorithms: Label Propagation and Louvain, specifically designed to leverage the capabilities of Arachne, which is a Python-accessible open-source framework for large-scale graph analysis. Our implementations achieve substantial speedups over existing Python-based tools like NetworkX and igraph, which lack efficient parallelization, and are competitive with parallel frameworks such as NetworKit. Experimental results show that Arachne-based methods outperform these baselines, achieving speedups of up to 710x over NetworkX, 75x over igraph, and 12x over NetworKit. Additionally, we analyze the scalability of our implementation under varying thread counts, demonstrating how different phases contribute to overall performance gains of the parallel Louvain algorithm. Arachne, including our community detection implementation, is open-source and available at \url{https://github.com/Bears-R-Us/arkouda-njit}.
\end{abstract}


\section{Introduction}
\label{s:introduction}
Community detection, also referred to as graph clustering, aims to partition the vertices of a network into groups characterized by dense intra-cluster connections and sparse inter-cluster connections. It is important to comprehend the structure of complex networks, as community members usually exhibit similar roles within the network. In recent years, community detection has had a wide range of applications in various fields, including urban transportation \cite{yildirimoglu2018identification}, social networks \cite{bedi2016community}, biology \cite{lewis2010function}, FinTech \cite{gavin2023community}, and more. 

As networks continue to grow in size, it is necessary to utilize effective parallel algorithms. For graphs with millions to billions of edges, only (near) linear-time community-detection algorithms are practical. To address this, we focus on two parallel algorithms. Label propagation \cite{raghavan2007near} is a simple yet effective method in which vertices iteratively adopt the most frequent community label among their neighbors until stable communities emerge. The Louvain algorithm \cite{blondel2008fast}, widely recognized for its heuristic modularity optimization, has proven highly efficient for clustering large, real-world networks. Its performance has been further improved through various optimizations and parallel implementations \cite{que2015scalable, ghosh2018distributed, lu2015parallel, shi2021scalable, sahu2023gve}. In this paper, we implement these two algorithms in parallel and evaluate them. Furthermore, they are integrated into the Arachne open-source framework \cite{rodriguez2022arachne}, allowing scalable exploratory graph analysis.

Arachne \cite{rodriguez2022arachne} is an actively developed framework, integrated with Arkouda \cite{merrill2019arkouda}, which combines the productivity of Python at the front end with the high-performance computing capability of the Chapel programming language \cite{chamberlain2007parallel} at the back end. Arachne provides a comprehensive set of graph analytics algorithms, organized within a cohesive framework, with all Chapel and Python code available on GitHub. Our research focuses on developing and integrating parallel graph algorithms into Arachne, enabling data scientists to take advantage of Python and use their laptops to perform real-world graph analysis on large computing platforms productively. 

The primary contributions of this paper are as follows:

1) \textbf{Scalable Parallel Implementations of Community Detection Algorithms:} We develop high-performance parallel versions of two widely used community detection algorithms: Label propagation and Louvain, optimized for large-scale graphs. Using the parallel computing capabilities of Chapel, our implementations efficiently process graphs with billions of edges, achieving up to 710× speedup over NetworkX, 75× over igraph, and 12× over NetworKit. 

2) \textbf{Comprehensive Evaluation across Various Graphs:} We conduct extensive experiments to assess the efficiency, scalability, and output quality of our implementations. The results show that our methods consistently outperform widely used Python-based libraries across a variety of real-world graphs, while delivering comparable or better solution quality than both sequential and parallel baselines.

3) \textbf{Integration with the Arachne Graph Analytics Framework:} This work demonstrates how Arachne, a flexible and scalable open-source framework, facilitates the efficient implementation and execution of advanced community detection algorithms. By leveraging Arachne's design and capabilities, we enable seamless scalability and high performance, providing a practical tool for analyzing massive graphs. This integration highlights Arachne's potential as a robust platform for a broad range of graph-analysis tasks beyond community detection.  

\section{Preliminaries}
\label{sec:preliminaries}
\subsection{Notation}
Let $G(V, E, w)$ be an undirected, weighted graph, where $V$ represents the set of $n$ vertices, $E \subseteq \{\{u,v\} \mid u,v \in V\}$ denotes the set of $m$ edges. $N(v)$ denotes the neighbor of $v$. An edge $e$ is said to be incident to a vertex $v$ if $v$ is one of the two vertices that connects $e$. Loops, edges that connect a vertex to itself, are also allowed. The weight associated with an edge $e_{ij}$ is denoted by $w_{ij}$. In the case of an unweighted graph, we assume that all edges have unit weight, i.e., $w_{ij} = 1$. 
Following the convention of \cite{nguyen2021leiden}, the degree of a vertex is defined as the number of its neighbors:
\[
\deg(v) = |\{e \mid e \in E, v \in e \}|
\]
For the weighted degree, we sum the weights of all incident edges, defined as:
\[
\deg_{w}(v) = \sum_{e \in E \mid v \in e} w(e) + L_v
\]
where $L_v$ is the weight of the loop on $v$ if one exists and is 0 otherwise. Loops are counted twice, similar to how they are treated in directed graphs.
The weighted volume of a set of vertices $S$ is the sum of their weighted degrees:
\[
\textit{vol}_{w}(S) = \sum_{v \in S} \deg_{w}(v)
\]
For two disjoint subsets of vertices $A, B \subset V$, the weighted cut is defined as the total weight of all edges between $A$ and $B$:
\[
\textit{cut}_{w}(A,B)=\sum_{\{u,v\}\mid \{u,v\} \in E, u \in A, v \in B}w(\{u,v\})
\]
We use $cut_w(A)$ as shorthand for the cut between $A$ and the rest of the graph, that is, $cut_w(A, V \setminus A)$. Similarly, $cut_w(v, A)$ is used to represent $cut_w(\{v\}, A)$.

\subsection{Community Detection}
Community detection is a widely used technique in network science to partition a network into a set of disjoint communities $C = {c_1, \dots, c_n}$, where each community $c_i \subseteq V$, $\cup c_i = V$, and $c_i \cap c_j = \emptyset$. Although some methods allow for overlapping communities, in this paper we assume communities to be disjoint, so overlapping is not considered in this context. 

\subsection{Modularity}
Modularity, introduced by Newman and Girvan in 2004 \cite{newman2004finding}, is a widely used metric to evaluate the quality of communities identified by heuristic-based community detection algorithms. Communities with high modularity are characterized by dense internal connections and sparse connections between them. The modularity ranges from -0.5 to 1 \cite{brandes2007modularity}, with higher values indicating better community structures. As mentioned in \cite{hamann2021scalable}, modularity can be written as:
\[
\sum_{c \in C} \frac{\textit{vol}_w(c)-\textit{cut}_w(C)}{\textit{vol}_w(V)}-\frac{vol_w(C)^2}{vol_w(V)^2}
\]
Modularity can be changed by moving vertices between communities. A move consists of removing a vertex from its current community and assigning it to a different one. Let $A$ represent the current community of $v$ and $B$ the target community. Denote by $A^-$ and $B^-$ the communities $A$ and $B$ that exclude $v$, respectively. The modularity change, $\triangle{Q}{v \to B}$, resulting from such a move is given by:
\begin{multline}
\label{equation:deltaQ}
    \text{$\triangle{Q}_{v \to B} = 2*(\frac{\textit{cut}_w(v,B^-)-\textit{cut}_w(v,A^-)}{\textit{vol}_w(V)}-$}  \\
    \text{$\deg_w(V)*\frac{\textit{vol}_w(B^-)-\textit{vol}_w(A^-)}{\textit{vol}_w(V)^2})$}
\end{multline}

\subsection{Arachne}
Arachne is an extension to the existing library, Arkouda, which provides Pandas- and NumPy-like operations at a supercomputing scale. Arachne integrates into typical exploratory data analytics workflows within Arkouda by providing methods to automatically convert tabular data into graph structures, enabling large-scale graph operations and queries. It efficiently executes graph analytical kernels in shared and distributed memory, transferring data as needed between compute nodes and processors as required.

Currently, Arachne includes a variety of optimized graph kernels such as connected components \cite{du2023minimum}, subgraph isomorphism \cite{dindoostvf2}, and k-truss \cite{du2022high}, among others.

\section{Algorithms}
\label{sec:algorithms}
In this section, we describe two parallel variants of existing community detection algorithms, as well as the implementation details.
\subsection{Parallel Label Propagation}
\subsubsection{Algorithm}
The Label Propagation Algorithm (LPA), introduced by Raghavan et al.~\cite{raghavan2007near}, is a widely used method for community detection in large-scale networks due to its simplicity and scalability. The algorithm initializes each vertex with a unique label and iteratively updates vertex labels based on their neighbors. In each iteration, a vertex adopts the label with the highest aggregate edge weight in its neighborhood (for weighted graphs), effectively forming communities through local greedy updates.
The process continues until convergence, defined as the number of label updates falling below a user-defined threshold $\theta$. LPA has a near-linear time complexity; Each iteration takes \bigO{m} time, and the total running time is \bigO{L \cdot m}, where $L$ is the number of iterations. Although $L$ depends on the network structure rather than the size, LPA typically converges in only a few iterations in practice, despite the lack of formal guarantees.
\subsubsection{Implementation}
\input{algorithms/PLP}
Algorithm~\ref{alg:PLP} shows the pseudocode of our parallel label propagation implementation. The algorithm is parallelized by dividing the vertex set across threads, each operating independently on a shared label array (line 14). This asynchronous approach accelerates convergence, but can lead to race conditions: When evaluating the neighborhood of vertex $u$ (line 15), its neighbor $v$ could hold the previous label $C_{i-1}(v)$ or the current label $C_i(v)$. However, these race conditions are acceptable and even beneficial. In the original description of the algorithm~\cite{raghavan2007near}, the vertices are traversed in random order. We make this step optional in our parallel version, relying instead on the inherent randomization provided by thread execution.

To improve efficiency, we maintain a set of active vertices, $V_{active}$, which is initialized at line 5. A vertex becomes inactive if its label remains unchanged after an iteration, as implemented in lines 19 and 20, and is removed from the active set. It is reactivated only if one of its neighbors updates its label, as handled in line 25. This selective update strategy significantly reduces redundant computations and focuses computation on regions of the graph that are still undergoing change.
The algorithm terminates when the number of updated vertices falls below a user-defined threshold, which is checked in line 9, or when the predefined maximum number of iterations is reached, as specified in line 7. These termination criteria ensure efficient execution while preserving community quality, particularly in the presence of high-degree vertices that may otherwise delay convergence.

\subsection{Parallel Louvain Algorithm}
The Louvain algorithm, introduced by Blondel \emph{et al.} \cite{blondel2008fast}, is a multi-level greedy method designed to maximize modularity and identify high-quality, disjoint communities in large networks. The approach begins by assigning each vertex to its own community, resulting in an initial singleton partition, where each vertex is in a community by itself. In this way, every vertex is assigned a community ID equal to its vertex ID, an integer that uniquely identifies each vertex.

The algorithm proceeds through two iterative phases: the local-moving phase and the aggregation phase. During the local-moving phase, each vertex $v$ evaluates whether moving to a neighboring community would increase modularity and greedily joins the community that provides the highest gain. In the aggregation phase, all vertices within the same community are merged into a super-vertex, reducing the graph's size.

These two phases together form one iteration of the algorithm, which is repeated until no further improvement in modularity can be achieved.

In our parallel implementation of the Louvain algorithm, we go beyond the straightforward approach of parallelizing the main for-loop. Our primary optimizations take advantage of the features of Arachne and Arkouda to accelerate both the local-moving and aggregation phases of the algorithm. The specific improvements and detailed workings of our approach are described below.
\subsubsection{Local-moving Phase}
\input{algorithms/Local_Moving}
The pseudocode for the local-moving phase is shown in Alg.~\ref{arg:local-moving}.

At the beginning of each local-moving phase, since we calculate $\triangle Q$ using Eq.~\ref{equation:deltaQ}, we need to initialize two arrays: \textit{volVertex} and \textit{volCom}, representing the volume of vertices and the volume of the community. In line 4, each vertex is assigned its own community by setting the community ID to its vertex ID, effectively creating singleton communities. As a result, both \textit{volVertex} and \textit{volCom} are arrays initialized in lines 5-6 with lengths equal to the number of vertices, and their initial values are set to $w(v)$, the weight of the respective vertex.

The core of our local moving phase lies in the parallel vertex movement evaluation process starting from line 9. For each vertex $v$, the algorithm concurrently examines whether moving $v$ to one of its neighboring communities yields a positive $\triangle Q$. Specifically in lines 13-15, we compute the $\triangle Q$ in parallel for each neighboring community. After identifying a neighboring community that produces the maximum increase in $\triangle Q$ on line 16, the algorithm updates both the community membership and volume information. When a beneficial move is found, lines 17-19 show how we adjust the volumes by simultaneously decreasing $\textit{volCom}[c]$ and increasing $\textit{volCom}[d]$ by $\textit{volVertex}[v]$, followed by updating the vertex's community ID.

To optimize computation, line 7 introduces an atomic Boolean array, \textit{needCheck}, which tracks whether a vertex should be reconsidered for movement in the next iteration. The parallel processing in line 11 ensures that a vertex is only evaluated if it is marked for checking, which happens when either its community or that of one of its neighbors has changed. This selective processing significantly reduces unnecessary computations. The operation shown on line 25 updates the $needCheck$ array for the next iteration based on the changes made in the current round. As specified in line 26, the local-moving phase continues this parallel process until either no vertex changes its community or the maximum iteration count is reached.

We note that, similar to other parallel Louvain implementations, the final result is not strictly deterministic. Due to concurrent updates or variations in thread scheduling, the final community assignment and the achieved modularity may differ across runs. This nondeterminism is a common characteristic of parallel heuristics and does not affect the overall quality or scalability of the algorithm \cite{lu2015parallel}.

\subsubsection{Aggregation Phase}
In the implementation of the aggregation phase of the Louvain algorithm, we adopt a streamlined but robust strategy to iteratively remap and construct higher-level graphs. After determining the updated community assignments for each vertex at the start of the aggregation phase, the community IDs are first remapped to a contiguous range from 0 to the total number of unique communities. This ensures efficient indexing and facilitates the subsequent graph processing.

Subsequently, the edge list is traversed to update the endpoints of each edge according to the new community IDs. During this step, edge weights are recalculated based on the interactions between the communities represented by the endpoints. This updated edge list, now representing the coarsened graph, is transmitted back to the Python front-end.

The Arachne framework leverages Arkouda's \texttt{GroupBy} and \texttt{Broadcast} functionality during the graph reconstruction process. This critical step automatically merges edges with identical endpoints, aggregating their weights to produce the final edge weights for the coarsened graph. This integration of \texttt{GroupBy} and \texttt{Broadcast} not only simplifies the graph reconstruction, but also ensures computational efficiency by offloading the aggregation process to Arkouda's optimized back-end.

Following this, the algorithm alternates between the local-moving and aggregation phases in an iterative manner. During each iteration, the vertices are reassigned to communities in the local-moving phase based on modularity optimization, and the edge list is updated in the aggregation phase to reflect the new community assignments and then reconstruct a new graph. These steps are repeated until no further changes in vertex community assignments occur, representing convergence and the termination of the algorithm. The overall process is outlined in the following pseudo-code.
\input{algorithms/Louvain}

\section{Implementation and Experimental Setup}
\subsection{Datasets}
In this section, we present an experimental analysis using several real-world datasets, including \texttt{com-amazon}, \texttt{com-dblp}, \texttt{com-youtube}, \texttt{com-livejournal}, \texttt{as-skitter}, and \texttt{com-orkut}. These datasets are commonly used in network analysis studies and are publicly available at SNAP \cite{snapnets}. The number of vertices in these datasets ranges from 0.335 million to 4.00 million, while the number of edges spans from 0.926 million to 117 million. Detailed information about each dataset is presented in Table~\ref{table:dataset}
\input{table/dataset}

\subsection{Experimental configuration}
The experiments were conducted on an AMD EPYC 7763 server with dual sockets and 512GB of memory. Each socket contains 64 cores, providing a total of 128 cores. The system is equipped with 64KB L1 caches, 512KB L2 caches, and 32MB L3 caches. Arachne is set up to work in the client-server model, where the client, typically a Python script or Jupyter notebook, interacts with an Arkouda server running on a high-performance computing (HPC) system. The relevant code was compiled using GCC version 13.2.0. We used NetworKit v11.0, NetworkX v3.1, and igraph v0.11.8 as baselines. Unless otherwise stated, all running times exclude I/O and graph loading time and only include the algorithmic execution.

\section{Experiment Results}

\subsection{Comparing performance}
We evaluate the performance of our parallel implementations of two community detection algorithms: Label propagation and Louvain, using Arachne. These implementations are compared with established Python libraries, including igraph, NetworkX, and NetworKit. The running time and modularity for each algorithm are measured five times, and the average is reported to ensure accuracy and consistency.

\subsubsection{Label Propagation Algorithm}
\begin{figure}[htbp]
    \centering
    \includegraphics[width=0.5\textwidth]{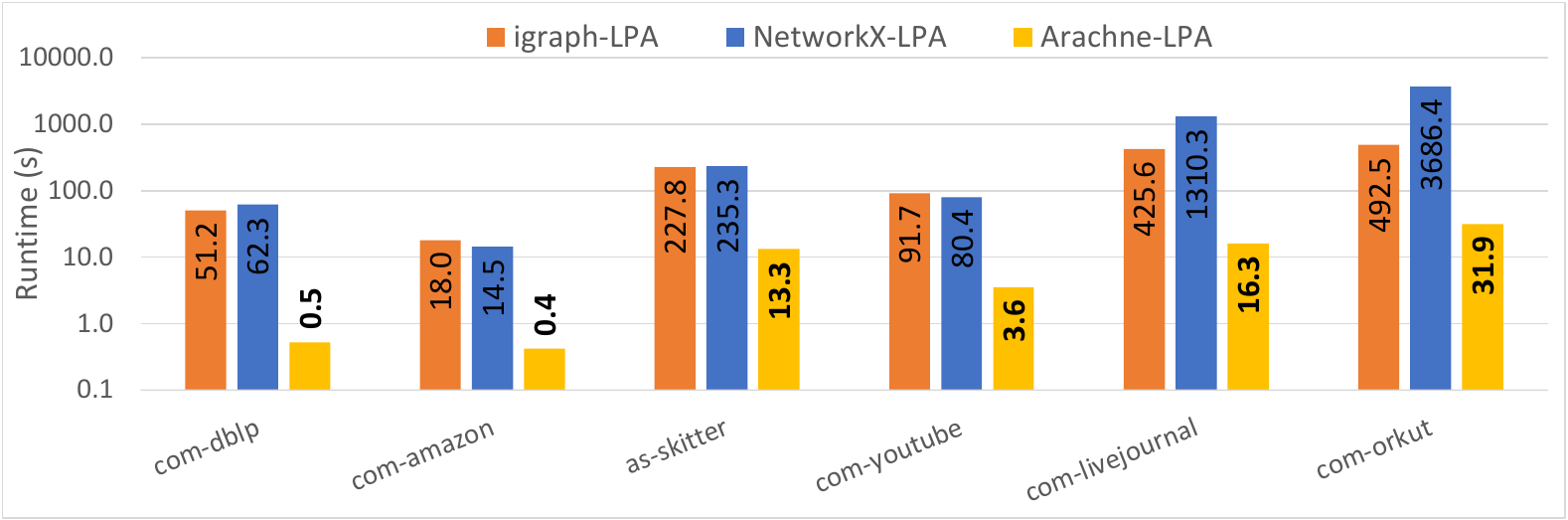}
    \caption{Running time in seconds with igraph-LPA, NetworkX-LPA, and Arachne-LPA}
    \label{fig:lpa_performance}
\end{figure}

As shown in Fig.~\ref{fig:lpa_performance}, our experimental results demonstrate that Arachne-LPA exhibits superior performance on all test graphs. For smaller graphs such as \texttt{com-dblp} and \texttt{com-amazon}, Arachne-LPA achieves execution times of 0.5s and 0.4s, respectively, demonstrating speedups of 96.9x and 42.7x compared to igraph implementations. This performance advantage becomes more pronounced with larger graphs: on \texttt{com-livejournal}, Arachne-LPA completes in 16.3s compared to igraph's 425.6s (26.2x speedup), while on \texttt{com-orkut}, it achieves a 15.4x speedup over igraph (31.9s vs 492.5s). In particular, NetworkX-LPA shows significant performance degradation on larger graphs, requiring 3686.4s for \texttt{com-orkut}.

\subsubsection{Louvain Algorithm}
\begin{figure}[htbp]
    \centering
        \includegraphics[width=0.5\textwidth]{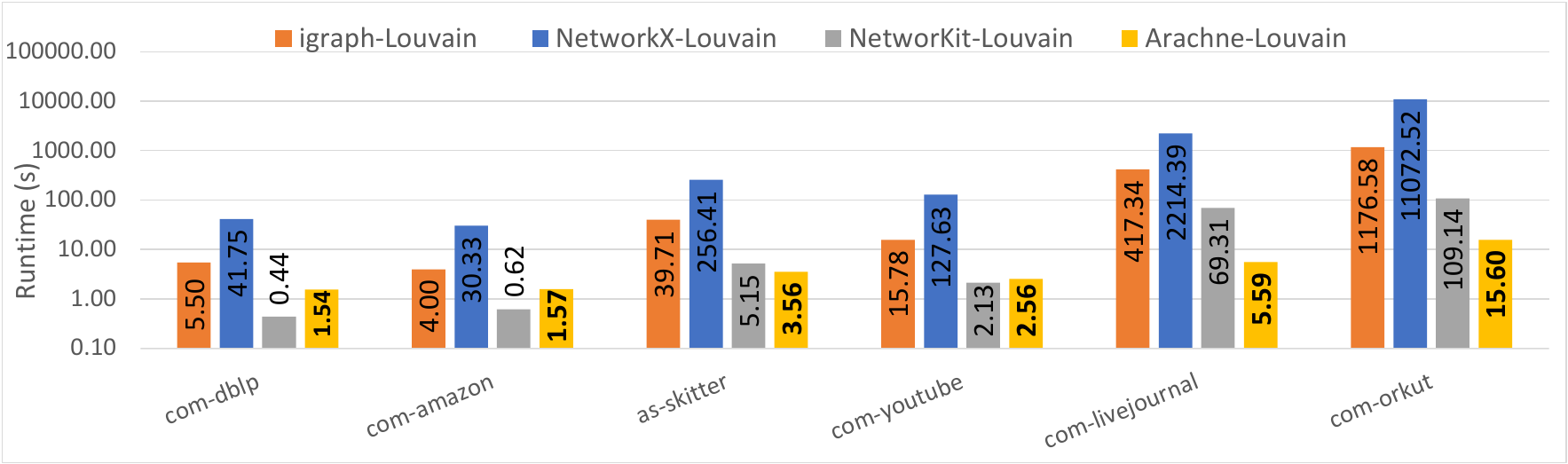}
        \caption{Running time in seconds with igraph-Louvain, NetworkX-Louvain, NetworKit-Louvain, and Arachne-Louvain}
        \label{fig:louvain_performance}
\end{figure}

\begin{figure}
    \centering
    \includegraphics[width=0.5\textwidth]{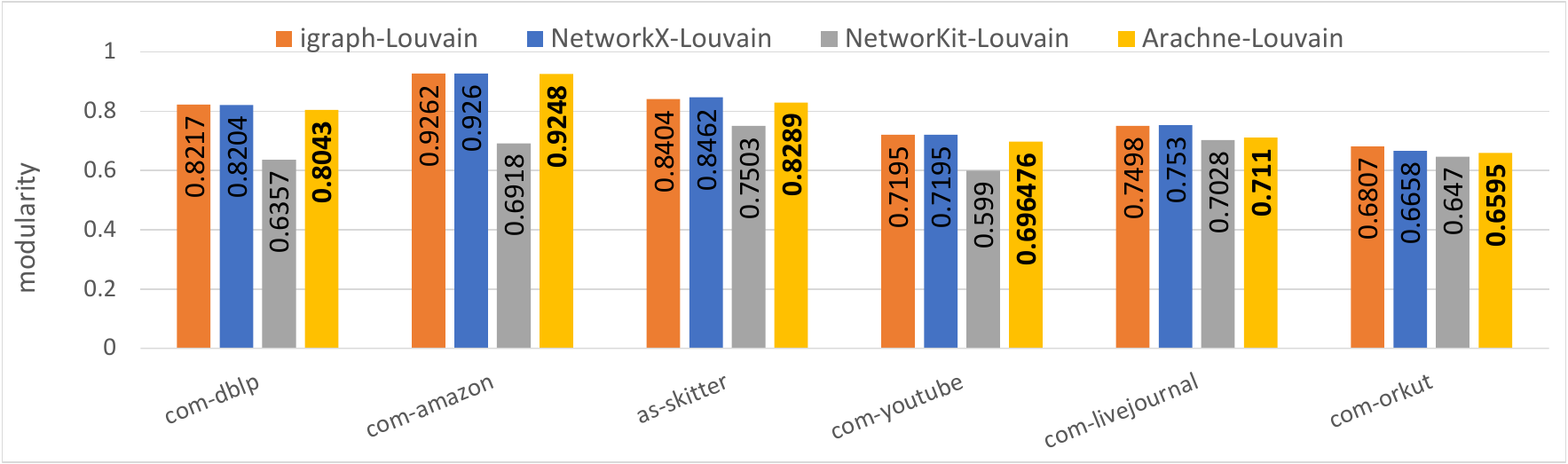}
    \caption{Modularity with igraph-Louvain, NetworkX-Louvain, NetworKit-Louvain, and Arachne-Louvain}
    \label{fig:louvain_modularity}
\end{figure}

The comparison of the running time of different Louvain implementations is presented in Fig.~\ref{fig:louvain_performance}. The Arachne-Louvain implementation consistently outperforms other methods in almost all the datasets tested. For smaller graphs like \texttt{com-dblp} and \texttt{com-amazon}, Arachne completes in 1.54s and 1.57s, respectively, compared to 5.50s and 4.00s for igraph, 41.75s and 30.33s for NetworkX, and 0.44s and 0.62s for NetworKit. Although NetworKit achieves lower running times on some small graphs, Arachne demonstrates strong scalability and becomes increasingly advantageous on larger datasets. For example, on \texttt{com-livejournal}, Arachne completes in 5.59s, significantly faster than igraph (417.34s), NetworkX (2214.39s), and NetworKit (69.31s). The greatest performance gain appears on \texttt{com-orkut}, where Arachne completes in 15.60s, compared to 1176.58s for igraph, 11072.52s for NetworkX, and 109.14s for NetworKit.
As shown in Fig.~\ref{fig:louvain_modularity}, Arachne-Louvain achieves modularity that is highly comparable to those produced by sequential implementations. For example, on \texttt{com-amazon} and \texttt{as-skitter}, Arachne reaches modularity values of 0.9248 and 0.8289, closely matching the results from igraph (0.9262 and 0.8404) and NetworkX (0.9260 and 0.8462). In contrast, NetworKit tends to produce lower modularity values on several datasets, such as 0.6918 on \texttt{com-amazon} and 0.7503 on \texttt{as-skitter}. These results suggest that Arachne not only provides significantly faster execution than sequential implementations but also delivers higher-quality results than existing parallel baselines like NetworKit.
\subsubsection{Analysis across Graph Sizes}
The experimental results reveal consistent performance characteristics between the two algorithms. Our experiments demonstrate that the relative performance advantage of Arachne implementations exhibits a strong correlation with the size of the graph. For smaller graphs such as \texttt{com-dblp} and \texttt{com-amazon}, we observe moderate but consistent performance improvements. However, efficiency becomes substantially more significant for large-scale graphs like \texttt{com-livejournal} and \texttt{com-orkut}, where performance improvements often span multiple orders of magnitude. This scaling behavior suggests that Arachne's parallel processing approach becomes particularly effective as graph size increases, making it especially suitable for large-scale graph analysis tasks.

\subsection{Strong Scaling}
\begin{figure}
    \centering
    \includegraphics[width=\linewidth]{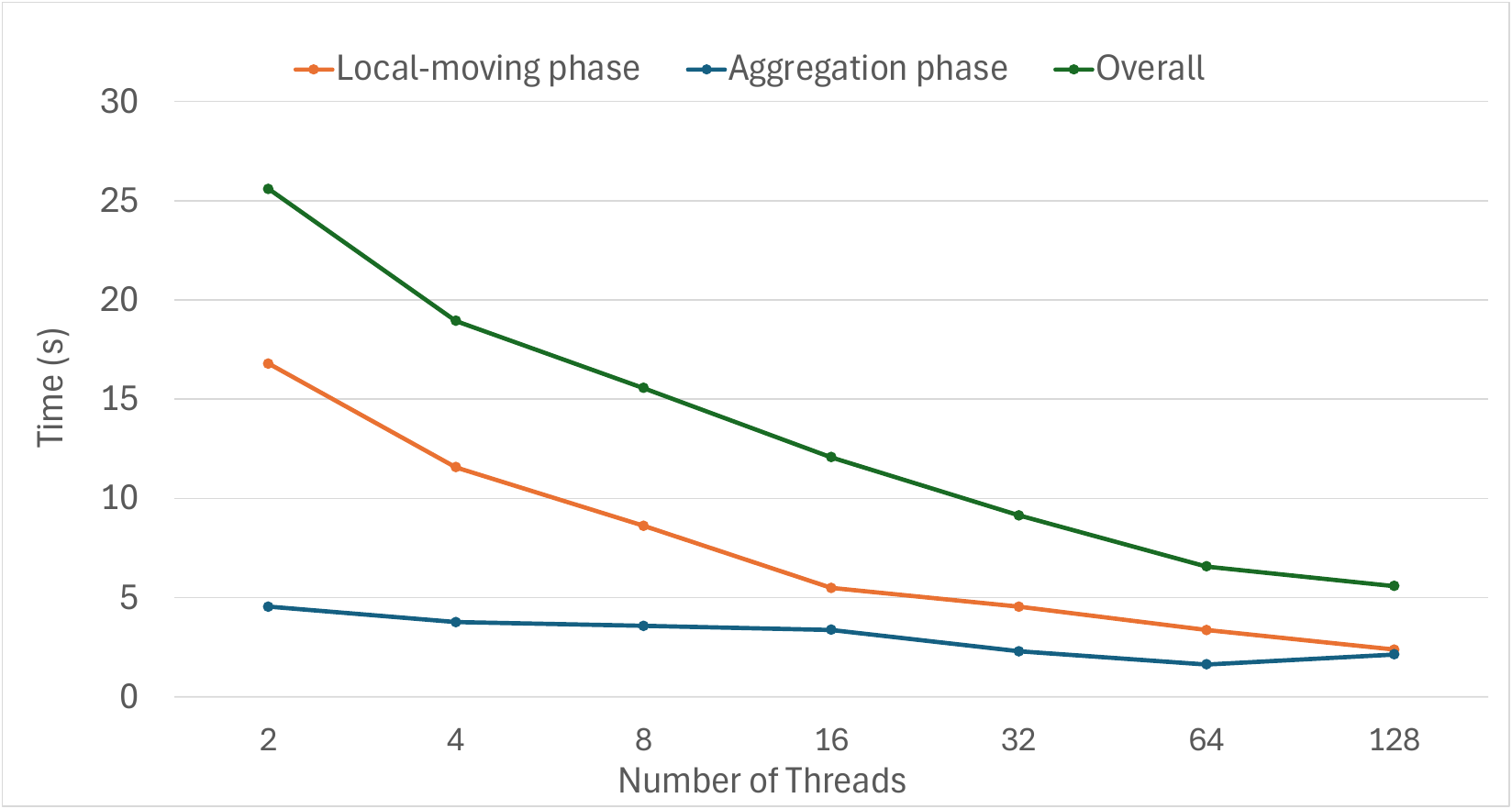}
    \caption{Scalability of execution times as the number of threads is increased for the Louvain implementation}
    \label{fig:louvain-scalability}
\end{figure}

Strong scaling experiments were conducted on the \texttt{com-livejournal} dataset to evaluate the parallel efficiency of our Louvain implementation, as shown in Fig.\ref{fig:louvain-scalability}. We varied the number of threads from 2 to 128 by configuring the \texttt{CHPL_RT_NUM_THREADS_PER_LOCALE} environment variable in Chapel. To provide deeper insights into the scaling behavior, we also break down the running time of different phases to better understand the individual performance. 

Our experimental results demonstrate that the Louvain implementation achieves strong scalability. As the number of threads increases, we observe substantial performance improvements. Specifically, the total execution time decreases from 25.6 seconds with 2 threads to 5.6 seconds with 128 threads, yielding a 4.6× speedup. 

When it comes to phase-level analysis, the local-moving phase, which dominates the sequential execution time in the Louvain algorithm, shows the most significant improvement with parallelization. Specifically, its execution time improves by 7x when scaling from 2 to 128 threads, reducing from 16.8s to 2.4s. This indicates that the label-update operations in this phase are highly parallelizable. In contrast, the aggregation phase exhibits limited scalability due to its global communication requirements.

\section{Related Work}
\label{sec:related_work}
%Community Detection has been extensively studied, with the label propagation \cite{raghavan2007near} and the Louvain algorithm \cite{blondel2008fast} being two of the most widely used heuristics.
Community detection has been extensively studied to address the problem from various perspectives, with label propagation \cite{raghavan2007near} and Louvain algorithm \cite{blondel2008fast} being two of the most widely used heuristics.

Label propagation has gained popularity due to its simplicity, efficiency, and ease of implementation. Various improvements have since been proposed to address its limitations \cite{wang2020label,zhang2014label}. The Louvain algorithm is a greedy modularity optimization method for detecting communities with high modularity. To scale it to large graphs, several parallel Louvain methods (PLMs) have been proposed \cite{lu2015parallel, naim2017community}. They demonstrated that parallel heuristics can achieve modularity comparable to sequential baselines while scaling to much larger graphs.
%A key challenge in PLMs is the synchronization of vertex updates. Some methods use non-real-time strategies where vertex states are updated only after each full iteration. For instance, Lu~\cite{lu2015parallel} implemented a non-real-time Louvain variant using OpenMP. In contrast, real-time strategies update vertex information immediately during local moves. Naim~\cite{naim2017community} adopts this approach using two hash tables to manage vertex and community data, achieving up to 270× speedup on medium-scale graphs. However, such methods can incur high storage and computation costs.

Several graph analysis frameworks expose these algorithms through Python interfaces. NetworkX \cite{hagberg2008exploring} is widely adopted in the Python ecosystem, but its pure Python implementation lacks parallelization. igraph \cite{csardi2006igraph} provides a C back-end with Python bindings, offering improved performance but only limited multithreading. NetworKit \cite{staudt2016networkit} is a high-performance C++ framework with Python bindings that includes parallel Louvain (PLM) and label propagation (PLP) implementations. Our comparisons focus on these frameworks because they represent the most widely used Python-facing tools, and thus provide the most relevant baselines to evaluate Arachne’s parallel performance.

Other state-of-the-art implementations, such as GPU-based algorithms \cite{forster2016louvain, sahu2025nu} or distributed-memory algorithms \cite{ghosh2018distributed}, target different execution environments. Although these achieve excellent performance on their respective platforms, they are beyond the scope of this work. 

%Other work focuses on leveraging hardware acceleration. Forster~\cite{forster2016louvain} employs CUDA to implement Louvain on GPUs, achieving up to 12× speedup compared to CPU-based versions. Bhowmick et al.\cite{bhowmik2019hydetect} proposed HyDetect, a hybrid CPU-GPU Louvain algorithm, and later extended it to support multi-node, multi-GPU systems\cite{bhowmick2022scalable}. Chou’s Nido system applies a batched clustering strategy optimized for GPUs, enabling processing of graphs exceeding aggregate GPU memory.

%Several graph analytics libraries provide community detection capabilities. NetworKit~\cite{staudt2016networkit} is a hybrid C++/Python framework offering parallel implementations of Louvain and Leiden~\cite{staudt2015engineering}. cuGraph, part of NVIDIA’s RAPIDS suite, delivers GPU-accelerated graph analytics through a Python interface, with core routines written in CUDA/C++.

\section{Conclusion and Future Work}
\label{sec:conclusion}
Large graph analysis presents significant challenges for data scientists, particularly in community detection, which is a common task. In this paper, we have developed and implemented several parallel and optimized algorithms for community detection with the open-source framework Arachne. Arachne facilitates large-scale graph analytics by leveraging supercomputers and cloud resources while providing a user-friendly Python interface. Our experimental results show that our implementation within the Arachne framework achieves substantial performance improvements, with absolute speedups compared to existing Python toolkits such as NetworkX, igraph, and NetworKit, demonstrating great scalability as well. 

For future work, we plan to expand the capabilities of Arachne by implementing additional community detection algorithms. In the meantime, our aim is to conduct comparative experiments against other state-of-the-art implementations to further validate the performance advantages of Arachne. By continuing to improve and optimize our software, we try to empower data scientists to analyze and derive insights from large-scale graph data.

\section{Acknowledgement}
\label{sec:acknowledgement}
We thank the Chapel and Arkouda communities for their support and guidance. This research is supported in part by NSF grants CCF-2109988, OAC-2402560, and CCF-2453324.

\bibliographystyle{IEEEtran}
\bibliography{ref}

\end{document}

%% file: algorithms/PLP.tex
\begin{algorithm}
\caption{Parallel Label Propagation (PLP).}
\label{alg:PLP}
\begin{algorithmic}[1]
\State \textbf{Input:} Graph $G = (V, E)$, $maxIteration$, $threshold$
\State \textbf{Output:} $C$: Community of each vertex

\vspace{1em} 
\Function{PLP}{$G$}
    \State $C \leftarrow [0..|V|]$ \Comment{initialized as singleton}
    \State $V_{\textit{active}} \leftarrow V$ \Comment{active vertices in each iteration}
    \State $\Delta N \leftarrow 0$ \Comment{number of vertices changed}
    \For{$l_i \in [0..\text{maxIteration}]$}
        \State $\Delta N \leftarrow \textsc{PLPMove}(G, C, V_{\textit{active}})$
        \If{$\Delta N  \leq \textsc{threshold}$}
            \State \textbf{break} 
        \EndIf
    \EndFor
    \State \Return $C$
\EndFunction

\vspace{1em} % Add blank space between functions

\Function{PLPMove}{$G, C, V_{\textit{active}}$}
    \State $\Delta N \gets 0$
    \State \textbf{parallel for} $v \in V_{\textit{active}}$ \textbf{do}
        \State \hspace{0.5cm} $c^* \leftarrow \textit{argmax}_{c}\{\sum_{u \in N(v):C(u)=c}w(v,u)\}$
        \State \hspace{0.5cm} \textbf{if} $c^* = C[v]$ \textbf{then}
            \State \hspace{1cm} $V_{\textit{active}} \leftarrow V_{\textit{active}} \setminus {v}$
            \State \hspace{1cm} \textbf{continue}
        \State \hspace{0.5cm} \textbf{end if}
        \State \hspace{0.5cm} $C[v] \leftarrow c^*$ 
        \State \hspace{0.5cm} $\Delta N \leftarrow \Delta N + 1$
        \State \hspace{0.5cm} $V_{\textit{active}} \leftarrow V_{\textit{active}} \cup N(v)$
    \State \textbf{end parallel for}
    \State \Return $\Delta N$
\EndFunction

\end{algorithmic}
\end{algorithm}

%% file: algorithms/Local_Moving.tex
\begin{algorithm}
\caption{Parallel Local-Moving Phase}
\label{arg:local-moving}
\begin{algorithmic}[1]
\State \textbf{Input:} Graph $G = (V, E)$, $maxIteration$
\State \textbf{Output:} Updated communities for all vertices

\For{$v \in V$}
    \State $comID[v] \leftarrow vertexID[v]$ \Comment{communityID}
    \State $volVertex[v] \leftarrow w(v)$ \Comment{volume of vertex}
    \State $volCom[v] \leftarrow w(v)$ \Comment{volume of community}
    \State $needCheck[v] \leftarrow true$
\EndFor
\State \textbf{repeat}
\State \hspace{0.5cm} \textbf{parallel for} $v \in V$ \textbf{do}
    \State \hspace{1.0cm} $tmpNeedCheck[v] \leftarrow false$
    \State \hspace{1.0cm} \textbf{if} $needCheck[v] = true$ \textbf{then}
        \State \hspace{1.5cm} \textbf{parallel for} $u \in N(v)$ \textbf{do}
            \State \hspace{2cm} $\Delta Q_{u} \leftarrow \Delta Q_{v\rightarrow comID[u]}$
        \State \hspace{1.5cm} \textbf{end parallel for}
        \State \hspace{1.5cm} $\Delta Q_{final} \leftarrow argmax_{u \in N(v)}\Delta Q_{u}$
        \State \hspace{1.5cm} \textbf{if} $\Delta Q_{final} > 0$ \textbf{then}
            \State \hspace{2cm} $volCom[comID[v]] \text{-=} volVertex[v]$
            \State \hspace{2cm} $volCom[comID[final]]\text{+=}volVertex[v]$
            \State \hspace{2cm} $comID[v] \leftarrow comID[final]$
            \State \hspace{2cm} $tmpNeedCheck[v] \leftarrow true$
            \State \hspace{2cm} \textbf{parallel for} $u \in N(v)$ \textbf{do}
                \State \hspace{2.5cm} $tmpNeedCheck[u] \leftarrow true$
            \State \hspace{2cm} \textbf{end parallel for}
        \State \hspace{1.5cm} \textbf{end if}
    \State \hspace{1cm} \textbf{end if}
\State \hspace{0.5cm} \textbf{end parallel for}
\State \hspace{0.5cm} $needCheck \leftarrow tmpNeedCheck$
\State \textbf{until} No vertex changes community or iteration count
\State \hspace{0.7cm} exceeds $maxIteration$
\State \textbf{return} $comID$

\end{algorithmic}
\end{algorithm}

%% file: algorithms/Louvain.tex
\begin{algorithm}
\caption{Parallel Louvain Algorithm}
\label{alg:louvain}
\begin{algorithmic}[1]
\State \textbf{Input:} Graph $G = (V, E)$, maxIteration $\theta$
\State \textbf{Output:} Final communities for all vertices
\Function{Louvain}{Graph $G$}
    \State \textbf{repeat}
        %\State \hspace{0.5cm} $P \leftarrow \text{SingletonPartition}(G)$
        \State \hspace{0.5cm} $C \leftarrow \text{LocalMoving}(G, \theta)$
        \State \hspace{0.5cm} $done \leftarrow |\{c \mid c \in C\}| == |V(G)|$ \Comment{Convergence}
        \State \hspace{0.5cm} \textbf{if} not $done$ \textbf{then}
            \State \hspace{1cm} $G \leftarrow \text{Aggregation}(G, C)$
            
        \State \hspace{0.5cm} \textbf{end if}
    \State \textbf{until} $done$
    \State \textbf{return} $C$
\EndFunction

\Function{Aggregation}{Graph $G$, Partition $C$}
    \State $V \leftarrow C$ \Comment{Remapping}
    \State $E \leftarrow \{\{A,B\} \mid \{u,v\} \in E(G):$
    \State \hspace{2.2cm} $u \in A \in C, v \in B \in C\}$
    \State \textbf{return} Graph$(V,E)$
\EndFunction

\end{algorithmic}
\end{algorithm}

%% file: table/dataset.tex
\begin{table}[!htbp]
    \centering
    \caption{Dataset Statistics}
    \label{table:dataset}
    \begin{tabular}{c|ccc}
    \toprule   Graph  & $|V|$ & $|E|$ & Diameter  \\
    \midrule
    com-dblp & 317,080	& 1,049,866 & 21 \\
    com-amazon & 334,863 & 925,872 & 44 \\
    as-skitter & 1,696,415	&11,095,298 & 25\\
    com-youtube &1,134,890 &2,987,624 &  20 \\
    com-livejournal &3,997,962&	34,681,189 & 17 \\
    com-orkut & 3,072,441	&117,185,083 & 9\\
    \bottomrule 
    \end{tabular}
\end{table}

%% file: main.bbl
\begin{thebibliography}{10}
\providecommand{\url}[1]{#1}
\csname url@samestyle\endcsname
\providecommand{\newblock}{\relax}
\providecommand{\bibinfo}[2]{#2}
\providecommand{\BIBentrySTDinterwordspacing}{\spaceskip=0pt\relax}
\providecommand{\BIBentryALTinterwordstretchfactor}{4}
\providecommand{\BIBentryALTinterwordspacing}{\spaceskip=\fontdimen2\font plus
\BIBentryALTinterwordstretchfactor\fontdimen3\font minus
  \fontdimen4\font\relax}
\providecommand{\BIBforeignlanguage}[2]{{%
\expandafter\ifx\csname l@#1\endcsname\relax
\typeout{** WARNING: IEEEtran.bst: No hyphenation pattern has been}%
\typeout{** loaded for the language `#1'. Using the pattern for}%
\typeout{** the default language instead.}%
\else
\language=\csname l@#1\endcsname
\fi
#2}}
\providecommand{\BIBdecl}{\relax}
\BIBdecl

\bibitem{yildirimoglu2018identification}
M.~Yildirimoglu and J.~Kim, ``{Identification of communities in urban mobility
  networks using multi-layer graphs of network traffic},'' \emph{Transportation
  Research Part C: Emerging Technologies}, vol.~89, pp. 254--267, 2018.

\bibitem{bedi2016community}
P.~Bedi and C.~Sharma, ``{Community detection in social networks},''
  \emph{Wiley interdisciplinary reviews: Data mining and knowledge discovery},
  vol.~6, no.~3, pp. 115--135, 2016.

\bibitem{lewis2010function}
A.~C. Lewis, N.~S. Jones, M.~A. Porter, and C.~M. Deane, ``{The function of
  communities in protein interaction networks at multiple scales},'' \emph{BMC
  systems biology}, vol.~4, no.~1, pp. 1--14, 2010.

\bibitem{gavin2023community}
J.~Gavin and M.~Crane, ``{Community detection in cryptocurrencies with
  potential applications to portfolio diversification},'' in \emph{FinTech
  Research and Applications: Challenges and Opportunities}.\hskip 1em plus
  0.5em minus 0.4em\relax World Scientific, 2023, pp. 177--202.

\bibitem{raghavan2007near}
U.~N. Raghavan, R.~Albert, and S.~Kumara, ``{Near linear time algorithm to
  detect community structures in large-scale networks},'' \emph{Physical review
  E}, vol.~76, no.~3, p. 036106, 2007.

\bibitem{blondel2008fast}
V.~D. Blondel, J.-L. Guillaume, R.~Lambiotte, and E.~Lefebvre, ``{Fast
  unfolding of communities in large networks},'' \emph{Journal of statistical
  mechanics: theory and experiment}, vol. 2008, no.~10, p. P10008, 2008.

\bibitem{que2015scalable}
X.~Que, F.~Checconi, F.~Petrini, and J.~A. Gunnels, ``{Scalable community
  detection with the Louvain algorithm},'' in \emph{2015 IEEE International
  Parallel and Distributed Processing Symposium}.\hskip 1em plus 0.5em minus
  0.4em\relax IEEE, 2015, pp. 28--37.

\bibitem{ghosh2018distributed}
S.~Ghosh, M.~Halappanavar, A.~Tumeo, A.~Kalyanaraman, H.~Lu,
  D.~Chavarria-Miranda, A.~Khan, and A.~Gebremedhin, ``{Distributed Louvain
  algorithm for graph community detection},'' in \emph{2018 IEEE international
  parallel and distributed processing symposium (IPDPS)}.\hskip 1em plus 0.5em
  minus 0.4em\relax IEEE, 2018, pp. 885--895.

\bibitem{lu2015parallel}
H.~Lu, M.~Halappanavar, and A.~Kalyanaraman, ``{Parallel heuristics for
  scalable community detection},'' \emph{Parallel Computing}, vol.~47, pp.
  19--37, 2015.

\bibitem{shi2021scalable}
J.~Shi, L.~Dhulipala, D.~Eisenstat, J.~\L{}\u{a}cki, and V.~Mirrokni,
  ``Scalable community detection via parallel correlation clustering,''
  \emph{Proc. VLDB Endow.}, vol.~14, no.~11, p. 2305–2313, Jul. 2021.

\bibitem{sahu2023gve}
S.~Sahu, K.~Kothapalli, and D.~S. Banerjee, ``High-performance implementation
  of {Louvain} algorithm with representational optimizations,'' in
  \emph{Complex Networks {\&} Their Applications XIII}, H.~Cherifi,
  M.~Donduran, L.~M. Rocha, C.~Cherifi, and O.~Varol, Eds.\hskip 1em plus 0.5em
  minus 0.4em\relax Cham: Springer Nature Switzerland, 2025, pp. 127--139.

\bibitem{rodriguez2022arachne}
O.~A. Rodriguez, Z.~Du, J.~Patchett, F.~Li, and D.~A. Bader, ``{Arachne: an
  Arkouda package for large-scale graph analytics},'' in \emph{2022 IEEE High
  Performance Extreme Computing Conference (HPEC)}.\hskip 1em plus 0.5em minus
  0.4em\relax IEEE, 2022, pp. 1--7.

\bibitem{merrill2019arkouda}
M.~Merrill, W.~Reus, and T.~Neumann, ``Arkouda: interactive data exploration
  backed by {C}hapel,'' in \emph{Proceedings of the ACM SIGPLAN 6th on Chapel
  Implementers and Users Workshop}, 2019, pp. 28--28.

\bibitem{chamberlain2007parallel}
B.~L. Chamberlain, D.~Callahan, and H.~P. Zima, ``{Parallel programmability and
  the chapel language},'' \emph{The International Journal of High Performance
  Computing Applications}, vol.~21, no.~3, pp. 291--312, 2007.

\bibitem{nguyen2021leiden}
F.~Nguyen, ``{Leiden-based parallel community detection},'' \emph{Bachelor's
  Thesis. Karlsruhe Institute of Technology}, 2021.

\bibitem{newman2004finding}
M.~E. Newman and M.~Girvan, ``{Finding and evaluating community structure in
  networks},'' \emph{Physical review E}, vol.~69, no.~2, p. 026113, 2004.

\bibitem{brandes2007modularity}
U.~Brandes, D.~Delling, M.~Gaertler, R.~Gorke, M.~Hoefer, Z.~Nikoloski, and
  D.~Wagner, ``{On modularity clustering},'' \emph{IEEE Transactions on
  Knowledge and Data Engineering}, vol.~20, no.~2, pp. 172--188, 2007.

\bibitem{hamann2021scalable}
M.~Hamann, ``{Scalable Community Detection},'' Ph.D. dissertation,
  Dissertation, Karlsruhe, Karlsruher Institut f{\"u}r Technologie (KIT), 2020,
  2021.

\bibitem{du2023minimum}
Z.~Du, O.~A. Rodriguez, F.~Li, M.~Dindoost, and D.~Bader, ``{Minimum-Mapping
  based Connected Components Algorithm}.''\hskip 1em plus 0.5em minus
  0.4em\relax The 10th Annual Chapel Implementers and Users Workshop (CHIUW),
  2023.

\bibitem{dindoostvf2}
M.~Dindoost, O.~A. Rodriguez, S.~Bagchi, P.~Pauliuchenka, Z.~Du, and D.~A.
  Bader, ``{VF2-PS: Parallel and Scalable Subgraph Monomorphism in
  Arachne}.''\hskip 1em plus 0.5em minus 0.4em\relax 28th Annual IEEE High
  Performance Extreme Computing Conference (HPEC), 2024.

\bibitem{du2022high}
Z.~Du, J.~Patchett, O.~A. Rodriguez, F.~Li, and D.~A. Bader, ``High-performance
  truss analytics in {Arkouda},'' in \emph{2022 IEEE 29th International
  Conference on High Performance Computing, Data, and Analytics (HiPC)}.\hskip
  1em plus 0.5em minus 0.4em\relax IEEE, 2022, pp. 105--114.

\bibitem{snapnets}
J.~Leskovec and A.~Krevl, ``{SNAP Datasets}: {Stanford} large network dataset
  collection,'' \url{http://snap.stanford.edu/data}, Jun. 2014.

\bibitem{wang2020label}
T.~Wang, S.~Chen, X.~Wang, and J.~Wang, ``{Label propagation algorithm based on
  node importance},'' \emph{Physica A: Statistical Mechanics and its
  Applications}, vol. 551, p. 124137, 2020.

\bibitem{zhang2014label}
X.-K. Zhang, X.~Tian, Y.-N. Li, and C.~Song, ``{Label propagation algorithm
  based on edge clustering coefficient for community detection in complex
  networks},'' \emph{International Journal of Modern Physics B}, vol.~28,
  no.~30, p. 1450216, 2014.

\bibitem{naim2017community}
M.~Naim, F.~Manne, M.~Halappanavar, and A.~Tumeo, ``{Community detection on the
  GPU},'' in \emph{2017 IEEE International Parallel and Distributed Processing
  Symposium (IPDPS)}.\hskip 1em plus 0.5em minus 0.4em\relax IEEE, 2017, pp.
  625--634.

\bibitem{hagberg2008exploring}
A.~Hagberg, P.~J. Swart, and D.~A. Schult, ``Exploring network structure,
  dynamics, and function using {NetworkX},'' Los Alamos National Laboratory
  (LANL), Los Alamos, NM (United States), Tech. Rep., 2008.

\bibitem{csardi2006igraph}
G.~Csardi and T.~Nepusz, ``The igraph software,'' \emph{Complex syst}, vol.
  1695, pp. 1--9, 2006.

\bibitem{staudt2016networkit}
C.~L. Staudt, A.~Sazonovs, and H.~Meyerhenke, ``{NetworKit: A tool suite for
  large-scale complex network analysis},'' \emph{Network Science}, vol.~4,
  no.~4, pp. 508--530, 2016.

\bibitem{forster2016louvain}
R.~Forster, ``{Louvain community detection with parallel heuristics on GPUs},''
  in \emph{2016 IEEE 20th Jubilee International Conference on Intelligent
  Engineering Systems (INES)}.\hskip 1em plus 0.5em minus 0.4em\relax IEEE,
  2016, pp. 227--232.

\bibitem{sahu2025nu}
S.~Sahu, N.~Mahen, and K.~Kothapalli, ``$\nu$-{LPA}: Fast {GPU}-based label
  propagation algorithm ({LPA}) for community detection,'' in \emph{2025 IEEE
  International Parallel and Distributed Processing Symposium Workshops
  (IPDPSW)}.\hskip 1em plus 0.5em minus 0.4em\relax IEEE, 2025, pp. 395--404.

\end{thebibliography}
